\documentclass[12pt]{iopart}
\usepackage[colorlinks=true,urlcolor=blue,linkcolor=blue]{hyperref}
\usepackage{graphicx}
\relax
\renewcommand{\theequation}{\arabic{section}.\arabic{equation}}

\def\be{\begin{equation}}
\def\ee{\end{equation}}
\def\bs{\begin{subequations}}
\def\es{\end{subequations}}
\def\calm{{\cal M}}
\def\calmb{{\bar{\cal M}}}
\def\wb{{\bar{w}}}

\def\lx{\lambda}
\def\ex{\epsilon}

\newcommand{\sx}{\sigma}

\newcommand{\Ab}{{\bar A}}
\newcommand{\Omb}{{\bar \Omega}}
\newcommand{\Hb}{{\bar H}}
\newcommand{\rb}{{\bar r}}

\newcommand{\lb}{{\bar \lx}}

\newcommand{\tb}{{\bar t}}
\newcommand{\ttb}{{\bar t}}
\newcommand{\fb}{{\bar f}}

\newcommand{\ct}{{\tilde c}}
\newcommand{\Rb}{{\bar R}}

\newcommand{\rt}{\tilde r}

\newcommand{\dw}{\delta w}

\newcommand{\rhb}{\bar{\rho}}

\newcommand{\kb}{\bar{k}}

\def\be{\begin{equation}}
\def\ee{\end{equation}}
\def\bs{\begin{subequations}}
\def\es{\end{subequations}}
\newcommand{\een}{\end{subequations}}
\newcommand{\ben}{\begin{subequations}}
\newcommand{\beq}{\begin{eqalignno}}
\newcommand{\eeq}{\end{eqalignno}}

\def \lta {\mathrel{\vcenter
     {\hbox{$<$}\nointerlineskip\hbox{$\sim$}}}}

\def\ms{M_{\rm s}}

\def\a{\alpha}

\newcommand\fverb{\setbox\pippobox=\hbox\bgroup\verb}
\newcommand\fverbdo{\egroup\medskip\noindent%
                        \fbox{\unhbox\pippobox}\ }
\newcommand\fverbit{\egroup\item[\fbox{\unhbox\pippobox}]}
\newbox\pippobox

\def \lta {\mathrel{\vcenter
     {\hbox{$<$}\nointerlineskip\hbox{$\sim$}}}}

\newcommand{\bea}{\begin{eqnarray}}
\newcommand{\bdm}{\begin{displaymath}}
\newcommand{\edm}{\end{displaymath}}

\newcommand{\eea}{\end{eqnarray}}
\def\sc{\bar{H}_i }
\newcommand{\bt}{\bar{t}}
\newcommand{\bR}{\bar{R}}
\newcommand{\brr}{\bar{r}}
\newcommand{\sbA}{\sqrt{\bar{A}}}
\newcommand{\sbAz}{\sqrt{\bar{A}^{(0)}}}
\newcommand{\sbAf}{\sqrt{\bar{A}^{(1)}}}
\newcommand{\sbAs}{\sqrt{\bar{A}^{(2)}}}

\begin{document}

\title{Light Propagation and Large-Scale Inhomogeneities}

\author{Nikolaos Brouzakis, Nikolaos Tetradis and Eleftheria Tzavara}
\address{
University of Athens, Department of Physics, University Campus,
Zographou 157 84, Athens, Greece}

\begin{abstract}
We consider the effect on the propagation of light
of inhomogeneities with sizes of order 10 Mpc or larger.
The Universe is approximated through a variation of the Swiss-cheese model.
The spherical inhomogeneities are void-like, with central
underdensities surrounded by compensating overdense shells. We study the propagation of
light in this background, assuming that the source and the observer occupy random
positions, so that each beam
travels through several inhomogeneities at random angles.
The distribution of luminosity distances for sources
with the same redshift is asymmetric, with a peak at a value larger than the
average one. The width of the distribution
and the location of the maximum increase with increasing
redshift and length scale of the inhomogeneities. We compute the induced dispersion and
bias on cosmological parameters derived from the supernova data.
They are too small to explain the perceived acceleration without dark energy,
even when the
length scale of the inhomogeneities is comparable to the horizon distance.
Moreover, the dispersion and bias induced by gravitational lensing
at the scales of galaxies or clusters of galaxies are larger by at least an order of
magnitude.
\end{abstract}
\maketitle

\section{Introduction}
\setcounter{equation}{0}

The observed deviation from homogeneity of the structure of the Universe
at small length scales poses the question of whether
the use of the Friedmann-Robertson-Walker (FRW) metric
is adequate for the discussion of the cosmological expansion.
The argument for the applicability of a homogeneous solution is based on the
observation that the matter distribution is homogeneous when averaged over
length scales of ${\cal O}(100)\,h^{-1}$ Mpc. On the other hand,
the conclusion drawn from observations of distant supernova \cite{accel1,accel2},
or the perturbations
of the cosmic microwave background (CMB) \cite{wmap}, that the recent cosmological
expansion is accelerating has placed the effect of structure formation on the
overall expansion under scrutiny.

We are interested in the influence of inhomogeneities
with sub-horizon characteristic scales on the perceived expansion
\cite{rasanenn,extra,barausse}.
The observed inhomogeneities in the matter distribution with length scales of
${\cal O} (10)\, h^{-1}$ Mpc or smaller
are large, so that the Universe cannot be approximated
as homogeneous at these scales.
It is important to have a quantitative estimate of the influence
of these inhomogeneities on the data used for the determination of the expansion rate.
As all the observations involve the detection of light signals, it is crucial
to understand the effect of the inhomogeneous background on light propagation.

The transmission of a light beam in a general gravitational background can be studied
through the Sachs optical equations \cite{sachs}. These describe the expansion
and shear of the beam along its null trajectory. Apart from the case of
a FRW background, the optical equations have been derived by Kantowski
for a Scharzschild background \cite{kantowski}.
He used them for the study of light transmission within the Swiss-cheese
model of the Universe \cite{swiss}. In this model the light propagates essentially
in empty space with a beam expansion larger than the
average. In rare instances it passes near a very dense clump of matter, which
produces significant shear and focusing of the beam \cite{kantowski2}.
Because of the randomness of such events, sources with the same intrinsic luminosity and
redshift may have different luminosity distances. The distribution
is peaked at a value of the luminosity distance larger than the one in
a homogeneous background \cite{holz,sereno}, even though the mean value remains unaffected.
If the sample of observed sources is small, it unlikely that a clump of matter will
be encountered during the beam propagation. The beams essentially propagate in empty
space, while the expansion is induced by the average energy density.
A phenomenological
equation has been proposed by Dyer and Roeder in order to
describe the case in which a fraction of the energy density is homogeneously distributed
while the remaining is in clumps that do not affect the light propagation \cite{dyer}.

The picture of the Universe we described above is applicable to length scales of
${\cal O}(1)\, h^{-1}$ Mpc or smaller, for which the dominant structures are
galaxies or clusters of galaxies. At larger distances the averaged matter distribution
has a smaller density contrast. The use of the Schwarzschild geometry for the description
of the inhomogeneities is not appropriate.
The Lemaitre-Tolman-Bondi (LTB) metric \cite{ltb} has
been employed often for the modelling of the Universe at scales of
${\cal O}(10)\, h^{-1}$ Mpc or larger
\cite{mustapha}--\cite{biswas}. Its use demonstrates that
inhomogeneities can induce deviations of the luminosity distance from
its value in a homogeneous background.
For example, it has been observed that
any form of the luminosity distance as a function of redshift can be
reproduced with the LTB metric \cite{mustapha}. This means that the supernova data
can be explained through an inhomogeneous matter distribution in the context of
this metric.
However, reproducing the
data requires a variation of the
density or the expansion rate over distances of
${\cal O}(100)\,h^{-1}$ Mpc or even larger \cite{alnes}--\cite{biswas}.
In order to avoid a conflict with the isotropy of the
CMB the location
of the observer must be within a distance of ${\cal O}(10)\,h^{-1}$ Mpc 
from the center of the spherical configuration
described by the LTB metric. In this sense, the 
explanation of the perceived acceleration relies on the
position of the observer. On the other hand, there are indications for the presence
of a very large void in our vicinity \cite{tomita,localvoid}.

Our work is based on the fundamental assumption that we do not occupy a special
position in the Universe. We are interested in determining the maximum
effect that large-scale inhomogeneities can have on the luminosity distance.
The density contrast of inhomogeneities with characteristics lengths of
${\cal O}(10)\, h^{-1}$ Mpc or larger is at most of ${\cal O}(1)$.
For this reason it seems unlikely that they can affect the propagation of light
more strongly than the inhomogeneities with lengths of
${\cal O}(1)\, h^{-1}$ Mpc or smaller, whose density contrast can be larger than 1 by
several orders of magnitude. On the other hand, the validity of the FRW metric is
questionable if inhomogeneities with lengths comparable to the horizon distance
$\sim 3\times 10^3\,h^{-1}$ Mpc
develop a density contrast of ${\cal O}(1)$. We shall allow for such a possibility in
order to obtain a quantitative
estimate of the effect on the luminosity distance.

In a previous publication \cite{brouzakis} we derived the optical equations for a general
LTB background \cite{ltb}.
We used them in order
to study light propagation in a variant of the Swiss-cheese model.
The inhomogeneities are modelled as spherical regions within which the geometry
is described by the LTB metric. At the boundary of
these regions the LTB metric is matched with the FRW metric
that describes the evolution in the region between the inhomogeneities.
The Universe consists of collapsing or
expanding inhomogeneous regions, while
a common scale factor exists that describes the expansion of the homogeneous
intermediate regions. This model is similar to the standard Swiss-cheese
model \cite{swiss}, with the replacement of the Schwarzschild metric with
the LTB one. For this reason we refer to it as the LTB Swiss-cheese model.

We focus on the effect of inhomogeneities on the luminosity
distance if the source and the observer do not occupy special positions in the Universe.
This is achieved by placing both the source and the observer within the homogeneous
region of the LTB Swiss-cheese model. During its path the light signal crosses
several inhomogeneous regions before reaching the observer. As we have mentioned,
similar studies \cite{holz,sereno} have discussed the influence of structures
such as galaxies or clusters of galaxies on light propagation. The beam shear plays
an important role in this effect, characterized as gravitational lensing \cite{lensing}.
We focus on
inhomogeneities of much larger length scale, of ${\cal O}(10)\, h^{-1}$ Mpc or larger.
The matter distribution, even though inhomogeneous, is more evenly distributed than
in the previous case. In particular, we assume that each spherical region has a
central underdensity surrounded by an overdense shell. The densities in the two
regions are comparable at
early times and differ by a factor ${\cal O}(1)$ during the later stages of the evolution.
The LTB Swiss-cheese model we construct in this way describes a Universe dominated by
spherical voids with the compensating matter concentrated in shells surrounding them.
The beam shear is negligible in our calculations, and the main effect arises from
the variations of the beam expansion because of the inhomogeneities.

In ref. \cite{brouzakis} we estimated the deviations of the luminosity distance
from its value in a homogeneous Universe by considering the extreme case in which
the light passes through the centers of all the inhomogeneities it encounters.
In this work we perform a more detailed statistical analysis, by considering light
beams with random impact parameters relative to the centers. We check that the
resulting luminosity distance as a function of the impact parameter is consistent
with flux conservation. For a given redshift we estimate the width of the distribution
of luminosity distances. We also determine the deviation of the maximum of the
distribution from the value in a homogeneous background. For a central underdensity
this value is positive and sets the scale for
the perceived increase in the expansion rate relative to a homogeneous Universe.
An analytical study with similarities with our work is described in ref. \cite{biswas2}.
It focuses, however, on the study of the redshift, which determines only partially the
luminosity distance.
No statistical analysis, which is the central point of our work, is performed either.

In the following section we summarize the geodesic and optical
equations in a LTB background. In section 3 we describe the cosmological evolution of
the inhomogeneous regions.
In section 4 we study, both numerically and analytically, the
effect of the inhomogeneity on the properties (redshift,beam area) of a light beam that
travels through it. In section 5 we
study the luminosity distance as a function of the angle at which the beam crosses
the inhomogeneity. We verify that the results are consistent with flux conservation.
In section 6 we consider multiple crossings of inhomogeneities by the
light beam and the effect on the luminosity distance. We calculate the
distribution of the deviations of the luminosity distance from its value
in a homogeneous background. In section 7
we estimate the effect on the determination of
cosmological parameters from supernova data.

\section{Luminosity distance in a Lemaitre-Tolman-Bondi (LTB) background}
\setcounter{equation}{0}

Under the assumption of spherical symmetry,
the most general metric for a
pressureless, inhomogeneous fluid is the
LTB metric \cite{ltb}.
It can be written in the form
\begin{equation}
ds^{2}=-dt^2+b^2(t,r)dr^2+R^2(t,r)d\Omega^2,
\label{metrictb}
\end{equation}
where $d\Omega^2$ is the metric on a two-sphere.
The function $b(t,r)$ is given by
\begin{equation}
b^2(t,r)=\frac{R'^2(t,r)}{1+f(r)},
\label{brttb} \end{equation}
where the prime denotes differentiation with respect to $r$, and
$f(r)$ is an arbitrary function.
The bulk energy momentum tensor has the form
\begin{equation}
T^A_{~B}={\rm diag} \left(-\rho(t,r),\, 0,\, 0,\, 0  \right).
\label{enmomtb} \end{equation}
The fluid consists of successive shells marked by $r$, whose
local density $\rho$ is time-dependent.
The function $R(t,r)$ describes the location of the shell marked by $r$
at the time $t$. Through an appropriate rescaling it can be chosen to satisfy
$R(0,r)=r$.

The Einstein equations reduce to
\begin{eqnarray}
\dot{R}^2(t,r)&=&\frac{1}{8\pi M^2}\frac{\calm (r)}{R}+f(r)
\label{tb1} \\
\calm'(r)&=&4\pi R^2 \rho \, R',
\label{tb2} \end{eqnarray}
where the dot denotes differentiation with respect to $t$, and
$G=\left( 16 \pi M^2 \right)^{-1}$.
The generalized mass function $\calm(r)$ of the fluid can be chosen
arbitrarily. Because of energy conservation
$\calm(r)$
is independent of $t$, while $\rho$ and $R$ depend on both $t$ and $r$.

Without loss of generality we consider
geodesic null curves on the plane with $\theta=\pi/2$.
The first geodesic equation is
\be
\frac{dk^0}{d\lx}+\frac{\dot{R}'R'}{1+f}\left(k^1\right)^2+
\dot{R}R\left( k^3\right)^2=0,
\label{gtb3}\ee
with $k^i={dx^i}/{d\lambda}$ and $\lambda$ an affine parameter along the
null beam trajectory. The second geodesic equation can be replaced by the
null condition
\begin{equation}
-\left(k^0\right)^2+\frac{R'^2}{1+f}\left(k^1\right)^2+R^2
\left( k^3\right)^2=0,
\label{gtb4}
\end{equation}
while the third one can be integrated to obtain
\be
k^3=\frac{c_{\phi}}{R^2}.
\label{cphi}
\ee

The equation for the beam area can be written as \cite{brouzakis}
\be
\frac{1}{\sqrt{A}}\frac{d^2\sqrt{A}}{d\lx^2}=
-\frac{1}{4M^2}\rho \left( k^0\right)^2  -\sigma^2.
\label{exx3}
\ee
The shear $\sx$ is generated by
inhomogeneities, for which the local energy density is different from
the average one \cite{brouzakis}.
It describes the deformations of the cross-section of the
beam, induced by the propagation within an inhomogeneous medium. The shear
is important when the beam passes near regions in which the
density exceeds the average one by several orders of magnitude.
Within our modelling of large-scale structure, applicable for scales above
${\cal O} (10)\, h^{-1}$ Mpc,
the average density contrast is not sufficiently large for the shear to become
important. (We have verified this conclusion numerically.)
For this reason we neglect it in our study.

The Friedmann-Robertson-Walker (FRW) metric is a special case of the LTB metric with
\begin{eqnarray}
&&R(t,r)=a(t)r
~~~~~~~~~~~~~~~
f(r)=cr^2,~~~c=0,\pm 1
\label{tbfrw1} \\
&&\rho=\frac{c_\rho}{a^{3}(t)}
~~~~~~~~~~~~~~~~~~~~~~~
\calm(r)=\frac{4\pi}{3} c_\rho r^3.
\label{tbfrw2} \end{eqnarray}
The geodesic equation (\ref{gtb3}) has the solution
\begin{equation}
k^0=\frac{c_t}{a(t)}.
\label{gsolfrw1}
\end{equation}
The solution of eq. (\ref{exx3}) for an outgoing
beam is
\be
A(\lx) = r^2(\lx)\, a^2(t(\lx)) \, \Omega_s.
\label{asol} \ee
If we normalize the scale factor so that $a(t_s)=1$ at the time
of the beam emission, we recover the standard expression
$A=r^2 \Omega_s$ in flat space-time.
The constant $\Omega_s$ can be identified with the solid angle spanned by a
certain beam when the light is emitted by a point-like isotropic source.
We are interested in light propagation in more general
backgrounds. We assume that the light emission near the source is not
affected by the large-scale geometry. By choosing an affine parameter
that is locally $\lx=t$ in the vicinity of the source, we can set
\be
\left. \frac{d\sqrt{A}}{d\lx} \right|_{\lx=0}=\sqrt{\Omega_s}.
\label{init1} \ee
This expression, along with
\be
\left. \sqrt{A} \right|_{\lx=0}=0,
\label{init2} \ee
provide the initial conditions for the solution of eq. (\ref{exx3}).

In order to define the luminosity distance, we consider photons
emitted within a solid angle $\Omega_s$
by an isotropic source with luminosity $L$.
These photons are detected
by an observer for whom the light beam
has a cross-section $A_o$.
The redshift factor is
\be
1+z=\frac{\omega_s}{\omega_o}=\frac{k^0_s}{k^0_o},
\label{redshiftt} \ee
because the frequencies measured at the source and at the observation point
are proportional to the values of $k^0$ at these points.
The energy flux $f_o$ measured by the observer is
\be
f_o=  \frac{L}{4\pi D_L^2}= \frac{L}{4\pi}
\frac{\Omega_s}{(1+z)^2 A_o}.
\label{lumm} \ee
The above expression allows the determination of the luminosity distance
$D_L$ as
a function of the redshift $z$. The beam area can be calculated by solving
eq. (\ref{exx3}), with initial conditions given by eqs.
(\ref{init1}), (\ref{init2}), while the redshift is given by
eq. (\ref{redshiftt}).

\section{Modelling the inhomogeneities}
\setcounter{equation}{0}

We study the effect of inhomogeneities on the luminosity
distance without assuming a preferred location of the observer.
We model the inhomogeneities as spherical regions within which the geometry
is described by the LTB metric. At the boundary of
these regions, the LTB metric is matched with the FRW metric
that describes the expansion of the homogeneous
intermediate regions. Our model is similar to the standard Swiss-cheese
model \cite{swiss}, with the replacement of the Schwarzschild metric with
the LTB one. For this reason we refer to it as the LTB Swiss-cheese model.

The choice of the two
arbitrary functions $\calm(r)$ and $f(r)$ in eq. (\ref{metrictb})
can lead to different physical situations.
The mass function $\calm(r)$ is related to the initial matter distribution.
The function $f(r)$ defines an effective curvature term
in eq. (\ref{tb1}).
We work in a gauge in which $R(0,r)=r$. We parametrize the initial
energy density as $\rho_i(r)=\left( 1+\ex(r)\right)\rho_{0,i}$, with
$\rho_i(r)=\rho(0,r)$ and $|\ex(r)| < 1$.
The initial energy density of the homogeneous background is
$\rho_{0,i}$. If the size of the inhomogeneity is
$r_0$, a consistent solution requires
$4\pi\int_0^{r_0} r^2 \ex(r) dr=0$,
so that
\be
\calm(r_0)=4\pi\int_0^{r_0}
r^2\rho(r)\, dr=\frac{4\pi}{3} r^3_0 \rho_{0,i}.
\label{req} \ee
This is obvious if we apply eqs.  
(\ref{tb1}), (\ref{tb2}) to the homogeneous part at $r > r_0$. The FRW metric
is a special case of the LTB metric, described by eqs. (\ref{tbfrw1}), (\ref{tbfrw2}). 
Consistency of the equations as the shell with $r=r_0$ is approached from both sides
imposes the condition (\ref{req}). Moreover, the absence of singularities requires 
the continuity of $f(r)$ and $f'(r)$. As we assume that the homogeneous part is flat,
this means that $f(r_0)=f'(r_0)=0$. 
Alternatively, one may consider the matching of the two metrics at the surface 
with $r=r_0$ employing junction conditions \cite{israel}. 
If this surface does not contain a singular energy density, 
the above constraints must be imposed \cite{biswas2,matching}.

We assume that
at the initial time $t_i=0$ the expansion rate $H_i=\dot{R}/R=\dot{R}'/R'$
is given for all $r$ by the standard
expression in homogeneous cosmology:
$H_i^2=\rho_{0,i}/(6M^2)$.
Then, eq. (\ref{tb1}) with $R(0,r)=r$ implies that
\be
f(r)=\frac{\rho_{0,i}}{6M^2}r^2\left(
1-\frac{3\calm(r)}{4\pi r^3 \rho_{0,i}}\right).
\label{fr} \ee
The spatial curvature of the LTB geometry is
\be
^{(3)}R(r,t)=-2 \frac{(fR)'}{R^2 R'}.
\label{spcurv} \ee
For our choice of $f(r)$ we find that at the initial time
\be
^{(3)}R(r,0)=-6 H^2_i
\left(1- \frac{\calm'}{4\pi  r^2 \rho_{0,i}} \right)
=-6H^2_i
\left( 1-\frac{\rho_i(r)}{\rho_{0,i}} \right).
\label{spcurvin} \ee
Overdense regions have positive spatial curvature, while underdense ones
negative curvature.
This is very similar to the initial condition considered
in the model of spherical collapse \cite{sphc1}.

When the inhomogeneity is denser near the center, we have
$f(r) < 0$ for $r < r_0$ and $f(r)=0$ for $r \geq r_0$.
It is then clear from eq. (\ref{tb1}) that, in an
expanding Universe, the central region will have $\dot{R}=0$ at some point in
its evolution and will stop expanding. Subsequently, it will reverse its
motion and start collapsing. The opposite happens if the inhomogeneity has a central
underdensity. In this case, the central region expands faster than the surrounding
denser spherical shell. The width of the shell
decreases, while its density increases. It is the latter configuration that is relevant
if we want to model the Universe as being composed mainly of voids separated by
thin dense regions.

Our expressions simplify if we switch to dimensionless variables.
We define $\ttb=t H_i$, $\rb=r/r_0$, $\Rb=R/r_0$, where
$H^2_i=\rho_{0,i}/(6M^2)$ is the initial homogeneous
expansion rate and $r_0$ gives the size of the inhomogeneity in
comoving coordinates.
The evolution equation becomes
\be
\frac{\dot{\Rb}^2}{\Rb^2}=\frac{3\calmb(\rb)}{4\pi\Rb^3}+
\frac{\fb(\rb)}{\Rb^2},
\label{eind} \ee
with $\bar{\calm}=\calm/(\rho_{0,i}r^3_0)$
and $\fb=6M^2f/(\rho_{0,i}r_0^2)=f/\Hb^2_i$, $\Hb_i=H_ir_0$.
The dot now denotes a derivative with respect to $\ttb$.

We take the affine parameter $\lx$ to have the dimension of time and
we define the dimensionless variables
$\lb=H_i\lx $,
$\kb^0=k^0$, $\kb^1=k^1/\Hb_i$, $\kb^3=r_0 k^3$.
The geodesic equations (\ref{gtb3})--(\ref{cphi}) maintain their form, with
the various quantities replaced by barred ones, and the combination
$1+f$ replaced by $\Hb_i^{-2}+\fb$. For geodesics going through
subhorizon perturbations with
$\Hb_i\ll 1$ the effective curvature term $\fb$ plays a minor role. However,
this term is always
important for the evolution of the perturbations, as can be seen from
eq. (\ref{eind}).
The optical equation
takes the form
\begin{equation}
\frac{1}{\sqrt{\Ab}}\frac{d^2\sqrt{\Ab}}{d\lb^2}=
-\frac{3}{2}\rhb \left( \kb^0\right)^2,
\label{exx3r}
\end{equation}
with $\rhb=\rho/\rho_{0,i}$. We omitted the shear, as it gives a negligible contribution
to our results.
The initial conditions (\ref{init1}), (\ref{init2}) become
\bea
&&\left. \frac{d\sqrt{\Ab}}{d\lb} \right|_{\lb=0}=\frac{1}{\Hb_i}\sqrt{\Omb_s}
=\sqrt{\Omega_s}.
\label{init1r} \\
&&\left. \sqrt{\Ab} \right|_{\lx=0}=0,
\label{init2r} \eea
with $\Ab=H_i^2A$ and $\Omb=\Hb_i^2 \Omega$.

\begin{figure}[t]
\includegraphics[width=12cm, angle=0]{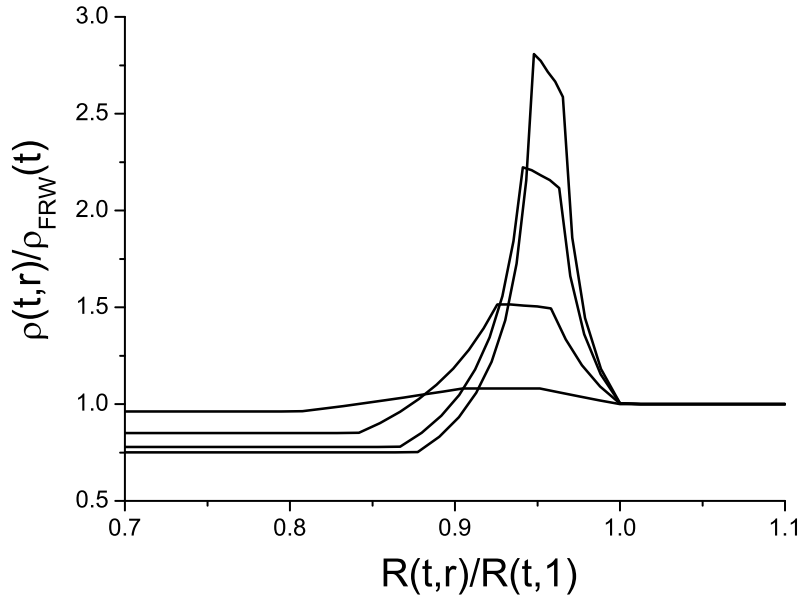}
 \caption{\it
The evolution of the density profile for a central underdensity surrounded by
an overdensity.
}
 \label{fig1}
 \end{figure}

\section{Single crossing}
\setcounter{equation}{0}

The typical cosmological evolution is displayed in fig.
\ref{fig1} for a central underdensity that is surrounded by an overdense region.
The initial density
$\rhb_i(r)=\rho_i(r)/\rho_{0,i}=1+\ex(r)$
is constant $\rhb_i=1+\ex_1$ in the region $\rb\leq 0.8$,
constant $\rhb_i=1+\ex_2$ in the region $0.9\leq \rb \leq 0.95$, and
$\rhb_i=1$ for $\rb \geq 1$. In the intervals $0.8 \leq \rb\leq 0.9$
and $0.95 \leq \rb \leq 1$ it interpolates linearly between the values
at the boundaries. This guarantees that the conditions $\fb(1)=\fb'(1)=0$, imposed by
the matching of the FRW and LTB metrics, are satisfied. 
We use $\ex_1=-0.01$. The value of $\ex_2$ is
fixed by the requirement that $\int_0^1\ex(\rb)\rb^2 d\rb=0$.
In a previous publication \cite{brouzakis}, we studied in detail the evolution 
of such inhomogeneities. 
We showed analytically that their growth is consistent with the standard theory of
structure formation and the spherical collapse model \cite{sphc1}. Inhomogeneities that
are initially of horizon size ($\Hb_i \sim 1$) with $|\ex_1| \sim 10^{-5}$ and 
have a characteristic scale 
${\cal O} (10)\, h^{-1}$ Mpc today also have a density contrast of ${\cal O}(1)$. 
In our numerical solutions we do not follow the very early evolution of the 
inhomogeneities. The perturbations we consider are already of subhorizon size and have 
a density contrast of ${\cal O}(10^{-2})$. They evolve to form present-day structures
with size ${\cal O} (10)\, h^{-1}$ Mpc or larger and density contrast of ${ \cal O}(1)$. 
The profile of the perturbations that we assume in this work differs slightly from
the one in ref. \cite{brouzakis} with respect to the width of the shell.
In ref. \cite{brouzakis} the widths of the underdense and overdense regions were taken 
equal. In this work we assume that the initial shells are
narrower and denser than in ref. \cite{brouzakis}. 
Clearly, our modelling of the Universe cannot reproduce all the details of the 
statistical nature of the inhomogeneities, especially during their evolution in the non-linear
regime. Our choice results in a model of the Universe dominated by voids, in rough agreement with
observations. It is possible, however, that the effect of underdensities is overestimated, 
as they occupy a slightly larger fraction of the total volume than the overdense regions already 
in the initial small perturbations.

In fig. \ref{fig1} we display the density profile at times $\ttb$=10, 100, 200, 250.
We follow the evolution at later times as well, even though we do not
depict it in fig. \ref{fig1}.
We normalize the energy density to that of a homogenous FRW background
(given by $\rhb_{FRW}(\tb)=\rhb(\tb,1)$). The earliest time corresponds to the
curve with the smallest deviation from 1, while the latest to the curve
with the largest deviation.
We observe that
the density contrast grows and eventually becomes of ${\cal O}(1)$.
The central energy density drops relative to the homogeneous background, while
the surrounding region becomes denser.
The radius of the central underdensity grows relative to the total size
of the inhomogeneity, as this region expands faster than the average.
The surrounding shell becomes thinner and denser.

\begin{figure}[t]
\includegraphics[width=12cm, angle=0]{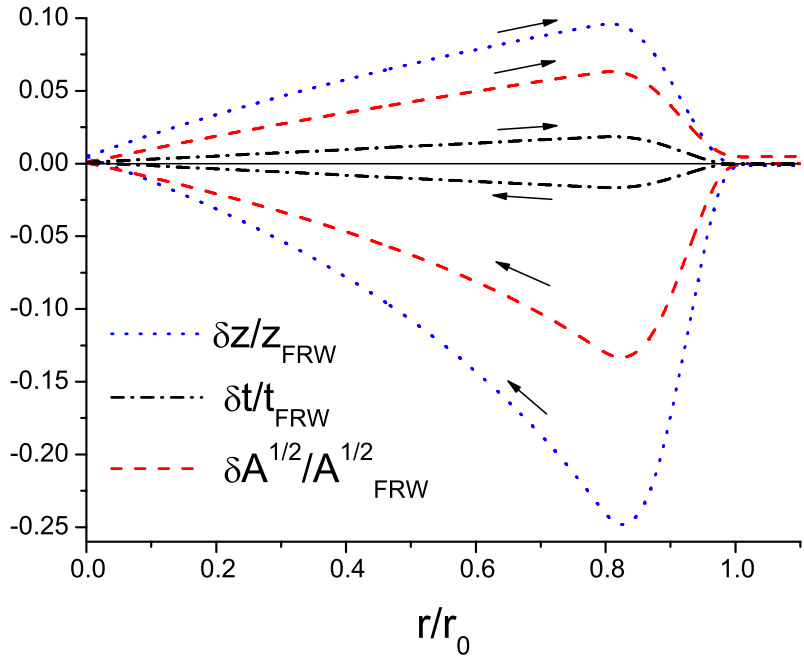}
 \caption{\it
The relative difference in redshift,
coordinate time, and
beam area, between the propagation in the
background of fig. \ref{fig1} and in a homogeneous background,
as a function of the radial coordinate.
}
 \label{fig2}
 \end{figure}

\begin{figure}[t]
\includegraphics[width=12cm, angle=0]{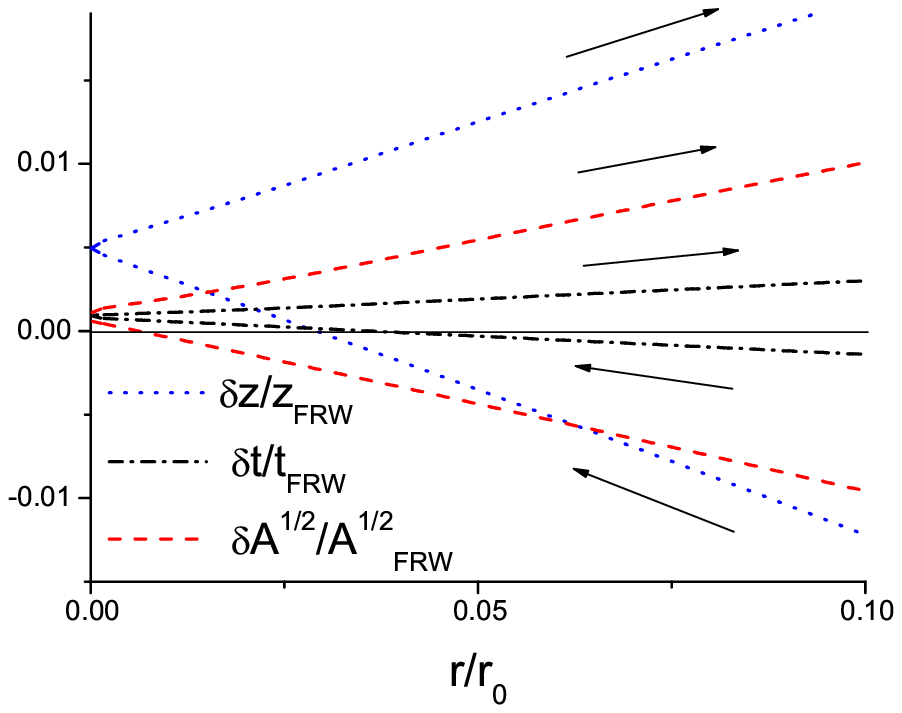}
 \caption{\it
The relative difference in redshift,
coordinate time, and
beam area, between the propagation in the
background of fig. \ref{fig1} and in a homogeneous background,
as a function of the radial coordinate.
}
 \label{fig3}
 \end{figure}

The effect of the spherical
inhomogeneity on the characteristics of a light beam is depicted
in figs. \ref{fig2}, \ref{fig3}.
(Fig. \ref{fig3} is a magnification of fig. \ref{fig2} around $\rb=0$.)
We assume a perturbation with $\Hb_i=1$ at the initial time $\ttb_i=0$, so that
the effects on the light beam are clearly visible. We use this perturbation for the 
numerical analysis both in this 
section and in section 5. Such a perturbation
leads to an inhomogeneity at the present time with size that would be in conflict with
observations (approximately $350\,h^{-1}$ Mpc). We consider realistic perturbations in section 6.
We consider a beam with $c_\phi=0$
that is emitted from a point with $\rb_s=1.5$ and passes through the center of symmetry.
The emission time is $\ttb_s = 250$ and the signal reaches $\rb=1.5$ again at a
time $\ttb\simeq 441$. The redshift for the exiting beam at $\rb=1.5$ is $z\simeq 0.46$.
We plot the relative difference in redshift $(z-z_{FRW})/z_{FRW}$,
coordinate time $(\ttb-\ttb_{FRW})/\ttb_{FRW}$, and
beam area  $\left( \sqrt{\Ab}-\sqrt{\Ab_{FRW}}\right)/\sqrt{\Ab_{FRW}}$, between the propagation in the
background of fig. \ref{fig1} and in a homogeneous background,
as a function of the radial coordinate $\rb$.
The arrows indicate the evolution of various quantities
as the beam enters the inhomogeneity from one side and exits from the other.
In the region
$0.1 \lta \rb \lta 0.9$
we observe a significant deviation of all quantities from their values in a
homogeneous background. However, the deviation becomes small in the region
near the origin. When the beam moves out of the inhomogeneous region only the
beam area deviates from the value in a homogeneous background. The coordinate
time and the redshift are not affected significantly.

We can obtain an understanding of the evolution depicted in figs. \ref{fig2}, \ref{fig3}
through an analytical treatment.
The effect of the inhomogeneity on the characteristics of the light beam can be
estimated analytically for perturbations with size much smaller than the distance to
the horizon. These have $\Hb_i \ll 1$.
Let us consider a beam with $c_\phi=0$ that passes through
the center of the spherical inhomogeneity.
The null condition (\ref{gtb4}) can be written as
\be
\frac{d\ttb}{d\rb}=\mp \Hb_i\frac{\Rb'}{\sqrt{1+\Hb^2_i \fb}}
\simeq \mp \Hb_i \Rb' \pm \frac{\Hb^3_i}{2} \Rb'\fb,
\label{null}\ee
for incoming and outgoing beams respectively.
If we keep terms up to ${\cal O}(\Hb^2_i)$ we can neglect the second
term in the above expression. This indicates that the spatial curvature does not
play a role if the inhomogeneities are much smaller than the horizon.
We can also employ the approximation
$\Rb'(\ttb,\rb)\simeq \Rb'(\ttb_s,\rb)+\dot{\Rb}'(\ttb_s,\rb)(\ttb-\ttb_s)$,
as the time it takes for the light to cross the inhomogeneity is much shorter than
the Hubble time. In fact, $\ttb-\ttb_s={\cal O}(\Hb_i)$
(see eqs. (\ref{sol1}), (\ref{sol2}) below).
We denote by $\rb_s$ the location of the source and by
$\ttb_s$ the emission time of the beam.
The solution of eq. (\ref{null}) is
\begin{eqnarray}
\ttb-\ttb_s=&&\Hb_i \left( \Rb(\ttb_s,\rb_s)-\Rb(\ttb_s,\rb)\right)
+\Hb_i^2\int^{\rb_s}_\rb \Rb'(\ttb_s,\rb) \dot{\Rb}(\ttb_s,\rb) d\rb
\nonumber \\
&&-\Hb_i^2 \left( \Rb(\ttb_s,\rb_s)-\Rb(\ttb_s,\rb)\right) \dot{\Rb}(\ttb_s,\rb)
+{\cal O}(\Hb_i^3)
\label{sol1}
\end{eqnarray}
for an incoming beam, and
\begin{eqnarray}
\ttb-\ttb_s=&&\Hb_i \left( \Rb(\ttb_s,\rb)-\Rb(\ttb_s,\rb_s)\right)
-\Hb_i^2\int_{\rb_s}^\rb \Rb'(\ttb_s,\rb) \dot{\Rb}(\ttb_s,\rb) d\rb
\nonumber \\
&&+\Hb_i^2 \left( \Rb(\ttb_s,\rb)-\Rb(\ttb_s,\rb_s)\right) \dot{\Rb}(\ttb_s,\rb)
+{\cal O}(\Hb_i^3)
\label{sol2}
\end{eqnarray}
for an outgoing one. These expressions are confirmed by numerical solutions.

We can also make a comparison with the propagation of light in a FRW background.
In this case we have $\Rb(\ttb,\rb)=a(\ttb)\rb=\Rb(\ttb,1)\rb$.
We have expressed the scale factor in terms of the value of the function $\Rb(\ttb,\rb)$
at the boundary of the inhomogeneous region $\rb=1$.
Let us consider light signals emitted at $\rb_s=1$ and observed at the center
($\rb_o=0$) of the inhomogeneity. The difference in propagation time within the
LTB and FRW backgrounds is
\begin{equation}
\ttb_o-\left(\ttb_o\right)_{FRW}=
\Hb_i^2\int_{0}^1 \Rb'(\ttb_s,\rb) \dot{\Rb}(\ttb_s,\rb) d\rb
-\frac{\Hb_i^2}{2} \Rb(\ttb_s,1) \dot{\Rb}(\ttb_s,1)
+{\cal O} (\Hb^3_i).
\label{dt1}
\end{equation}
For signals originating at $\rb_s=0$ and detected at $\rb_o=1$
the time difference has the opposite sign. As
a result, the time difference for signals that cross the
inhomogeneity is of ${\cal O} (\Hb^3_i)$. This is in agreement with figs. \ref{fig2},
\ref{fig3}.

We can derive similar expressions for the redshift of a light beam that
passes through the center of the inhomogeneity.
The geodesic equation (\ref{gtb3}) can be written as
\be
\frac{1}{k^0} \frac{dk^0}{d\rb}=
-\frac{d\,\ln(1+z)}{d\rb}=
=\pm \Hb_i\frac{\dot{\Rb}'}{1+\Hb^2_i \fb}
\simeq \pm \Hb_i\dot{\Rb}',
\label{redshift}\ee
for incoming and outgoing beams respectively.
In this way we find
\begin{eqnarray}
\ln(1+z)=&&\Hb_i \left( \dot{\Rb}(\ttb_s,\rb_s)-\dot{\Rb}(\ttb_s,\rb)\right)
\nonumber \\
&&-\Hb_i^2\int_{\rb_s}^\rb \ddot{\Rb}'(\ttb_s,\rb)
\left(\Rb(\ttb_s,\rb_s)-\Rb(\ttb_s,\rb) \right)d\rb
+{\cal O}(\Hb_i^3)
\label{solred1}
\end{eqnarray}
for an incoming beam, and
\begin{eqnarray}
\ln(1+z)=&&\Hb_i \left( \dot{\Rb}(\ttb_s,\rb)-\dot{\Rb}(\ttb_s,\rb_s)\right)
\nonumber \\
&&+\Hb_i^2\int_{\rb_s}^\rb \ddot{\Rb}'(\ttb_s,\rb)
\left(\Rb(\ttb_s,\rb)-\Rb(\ttb_s,\rb_s) \right)d\rb
+{\cal O}(\Hb_i^3)
\label{solred2}
\end{eqnarray}
for an outgoing one. These expressions are confirmed by numerical solutions.

For signals originating at $\rb_s=1$ and detected at $\rb_o=0$
the redshifts obey
\begin{eqnarray}
\ln \left(\frac{1+z}{1+z_{FRW}} \right)=
&&\Hb_i^2\int_0^1 \ddot{\Rb}'(\ttb_s,\rb)
\left(\Rb(\ttb_s,1)-\Rb(\ttb_s,\rb) \right)d\rb
\nonumber \\
&&-\frac{\Hb_i^2}{2} \ddot{\Rb}'(\ttb_s,1)\Rb(\ttb_s,1)
+{\cal O}(\Hb_i^3).
\label{relred} \end{eqnarray}
For signals originating at $\rb_s=0$ and detected at $\rb_o=1$
the r.h.s. of the above equation has the opposite sign. As
a result, the redshift difference for signals that cross the
inhomogeneity is of ${\cal O} (\Hb^3_i)$. Again, this is in agreement with figs. \ref{fig2},
\ref{fig3}.

\section{Luminosity distance and flux conservation}
\setcounter{equation}{0}

The beam area obeys the second-order differential equation (\ref{exx3r}),
whose solution depends crucially on the initial conditions. For initial conditions
given by eqs. (\ref{init1r}), (\ref{init2r}) and symmetric situations,
it is possible to determine the solution analytically. For signals emitted from some point $\rb_s$ at a
time $\ttb=\ttb_s$ and observed at $\rb_o=0$
we have \cite{partovi,celerier}
\begin{equation}
\sqrt{\Ab}= (1+z)\Rb(\ttb_s,\rb_s)\sqrt{\Omb}.
\label{sqrt1} \end{equation}
This is in agreement with figs. \ref{fig2},
\ref{fig3}.
For signals emitted from the center $\rb_s=0$
and observed at $\rb_o$ at a time $\ttb_o$
we have
\begin{equation}
\sqrt{\Ab}= \Rb(\ttb_o,\rb_o)\sqrt{\Omb}.
\label{sqrt2} \end{equation}
However, for a signal that crosses the inhomogeneity we need to integrate
eq. (\ref{exx3r}) from $\rb=0$ to $\rb_o$
with initial conditions determined by the propagation from $\rb_s$ to $\rb=0$. These
are different from (\ref{init1r}), (\ref{init2r}), so that an analytical
solution is not easy.

\begin{figure}[t]
\includegraphics[width=12cm, angle=0]{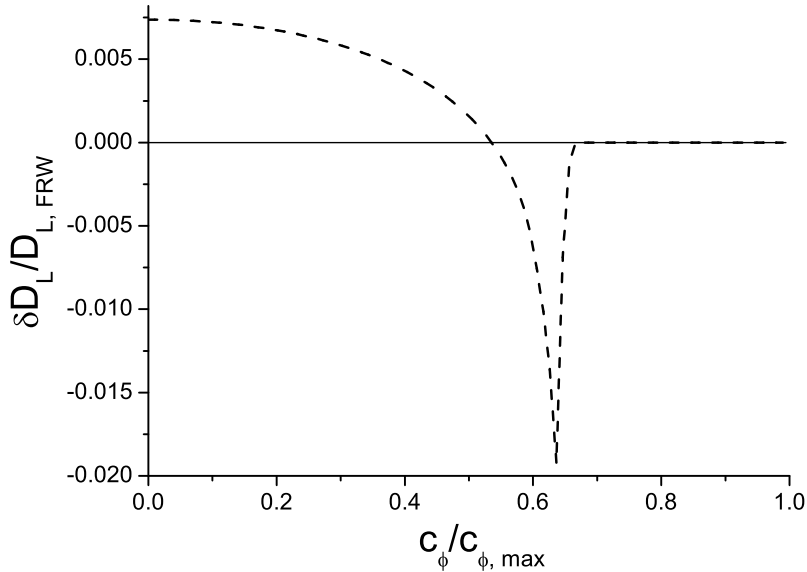}
 \caption{\it
The deviation of the luminosity distance from its value in a homogeneous background,
as a function of the impact parameter.
}
 \label{fig4}
 \end{figure}

Using perturbation theory, 
it is possible to obtain an analytical estimate 
of the deviation of the
luminosity distance from its value in homogeneous cosmology. 
We consider beam trajectories that start at the boundary of the inhomogeneity, 
pass through its center and exit from the other side.
The optical equation (\ref{exx3r}) can be written in the form 
\begin{equation}
\left( \kb^1\right)^2
\frac{d^2\sqrt{\Ab}}{d\rb^2}
+\frac{d\kb^1}{d\lb}
\frac{d\sqrt{\Ab}}{d\rb}
=
-\frac{3}{2}\rhb \left( \kb^0\right)^2 \sqrt{\Ab}.
\label{exx3rr}
\end{equation}
We express $d\bar{k}_1/d\bar{\lambda}$ in the above equation 
using the geodesic equation and
obtain
\be \frac{d^2\sbA}{d \brr^2}+\left(\pm \frac{2\sc
\dot{\bR}'}{\sqrt{1+\sc^2\fb}}-
\frac{\bR''}{\bR'}+\frac{\sc^2\fb'}{2(1+\sc^2 \fb)} \right) \frac{d
\sbA}{d \brr}=-\frac{3}{2} \bar{\rho} \frac{R'^2}{1+\sc^2 \fb} \sbA
\label{eq1} \ee 
where the positive sign in the second term corresponds to ingoing
and the negative sign to outgoing geodesics.

We use a simplified initial configuration for this estimate. We take $\bar{\rho}(0,\brr)=0$ for
$\brr<\brr_1$ and $\bar{\rho}(0,\brr)=1/(1-\brr_1^3)$ for $\brr>\brr_1$.
The initial conditions for an ingoing beam 
can be taken $\sbA(\rb=1)=0$, $d\sbA(\rb=1) /d\brr=-1$, without loss of generality.
We use the expansion 
\be
\sbA=\sbAz+\sc \sbAf+\sc^2 \sbAs +{\cal O}(\Hb_i^3),
\ee
and calculate $\sqrt{\Ab^{(i)}}$ in each order of perturbation theory. The 
calculation in described in the appendix. We point out that our choice of initial configuration, 
that involves
a discontinuous energy density, results in the appearance of $\delta$-function singularities 
in the second derivatives of $\Rb$ with respect to $\rb$. These must be taken into account in
a consistent calculation, as described in the appendix. We have checked that the expressions 
(\ref{sqrt1}) and (\ref{sqrt2}) are reproduced correctly by our results, up to second order in $\Hb_i$.

When the photon exits the inhomogeneity
at $\brr=1$ we
find
\begin{eqnarray}
\sbA(\brr=1)&=&2+4\sc+\left(5-\frac{3}{\brr_1^2+\brr_1+1}\right) \sc^2
+{\cal O}(\Hb^3_i)
\label{final} \\
\frac{d\sbA}{dr}(\brr=1)&=&1+4\sc+\left(6-\frac{3}{\brr_1^2+\brr_1+1}\right) \sc^2
+{\cal O}(\Hb^3_i).
\label{finalp} \end{eqnarray}
For a homogeneous universe we have
\begin{eqnarray}
\sbA(\brr=1)&=&2+4\sc+2\sc^2
\label{finalfrw} \\
\frac{d\sbA}{dr}(\brr=1)&=&1+4\sc+3\sc^2.
\label{finalpfrw} \end{eqnarray}
It is clear that the effect of the inhomogeneity is of
${\cal O}(\Hb_i^2)$. This conclusion has been confirmed by numerical solutions of the
exact optical equations, without any approximations. Several beam crossings
were studied for a multitude of values of $\Hb_i$. The 
coordinate time $\tb_o$ needed for the crossing, the redshift $z$ and the
beam area ${\Ab}$ were plotted in terms of $\Hb_i$. 
The polynomial fits to these plots verify with good precision the conclusions of
the present and previous sections:
The deviations of $\tb_o$ and $z$ from their values in a homogeneous background
are of ${\cal O}(\Hb_i^3)$, 
while the deviations of $\Ab$ are of ${\cal O}(\Hb_i^2)$.

From the above we can conclude that the main effect of an inhomogeneity is to
modify the luminosity distance of a light source behind it, while leaving
the redshift largely unaffected. In fig. \ref{fig4} we depict the modification
of the luminosity distance
$\left( D_L-\left(D_L \right)_{FRW}\right)/
\left(D_L \right)_{FRW}$
for light beams that cross a spherical inhomogeneity at
various angles. The inhomogeneity is the one employed in the
previous section. Its size is
approximately $350\,h^{-1}$ Mpc, much larger that what is deduced from observations.
We consider realistic inhomogeneities in the following section.
The crossing at a varying angle is achieved by choosing different values for the
constant $c_\phi$ in eq. (\ref{cphi}).
The light source is always located at the same point with $\rb_s=1.5$. The beam is
allowed to propagate until the redshift reaches a certain value $z_o\simeq 0.46$.
The beam with $c_\phi=0$ that goes through the center reaches $r_o\simeq 1.5$ at this
time.
The difference of the resulting luminosity distance from the one in a FRW background
is depicted in fig. \ref{fig4} as a function of $\ct_\phi=c_\phi/c_{\phi,max}$.
The largest value $c_{\phi,max}$ corresponds to beams that are emitted tangentially
with respect to the center of symmetry. The angle $\theta$ between the initial
direction of
the beam and the radial direction is determined through the relation $\sin\theta=\ct_\phi$.
The variable $\ct_\phi$ plays the role of an impact parameter, normalized to
1 for light beams emitted tangentially.
In fig. \ref{fig4} we observe that,
if $\ct_\phi$ is sufficiently small for the light to travel through
the central underdense region, the luminosity distance is enhanced relative to the
homogeneous case. For larger values of $\ct_\phi$ the light travels mainly through
the overdense shell and the luminosity distance in reduced.

If the redshift is not affected significantly by the propagation in the inhomogeneous
background, the conservation of the total flux implies that the average luminosity
distance must be the same as in the homogeneous case \cite{weinberg,rose}. As we
have seen, this is not
the case for an observer located at the center of an underdensity.
The resulting increase in the luminosity distance, arising mainly from the increase
in the redshift, can by employed for the explanation of the supernova data, even though
the size of the required inhomogeneity is probably in conflict with the observed large-scale
structure \cite{mustapha}--\cite{biswas}.
On the other hand, if the source and the observers are located outside the
inhomogeneity, as for fig. \ref{fig4}, so that the redshift is essentially unaffected,
we expect that the energy flux may be redistributed in various directions but the total
flux will be the same as in the homogeneous case \cite{weinberg,rose}.
This implies that the integral of $\left( D_L-\left(D_L \right)_{FRW}\right)/
\left(D_L \right)_{FRW}$ over all angles must vanish.
For an isotropic source
the integration over the solid angle is $2 \pi \sin\theta \, d\theta$.
This is equivalent to the integration over the
impact factor $\ct_\phi$, with a weight $\ct_\phi/\sqrt{1-\ct_\phi^2}$.
The integral of the function depicted in fig. \ref{fig4} is indeed approximately
zero. An analytical proof is very difficult, but a numerical analysis shows that a
 cancellation at the 90\% level takes place between the positive and
negative contributions. The remaining deviation from zero is caused by
numerical errors and the small (of ${\cal O}(\Hb_i^3)$), but non-vanishing, difference in redshifts
in inhomogeneous and homogeneous backgrounds.

\section{Multiple crossings}
\setcounter{equation}{0}

\begin{figure}[t]
\includegraphics[width=14cm, angle=0]{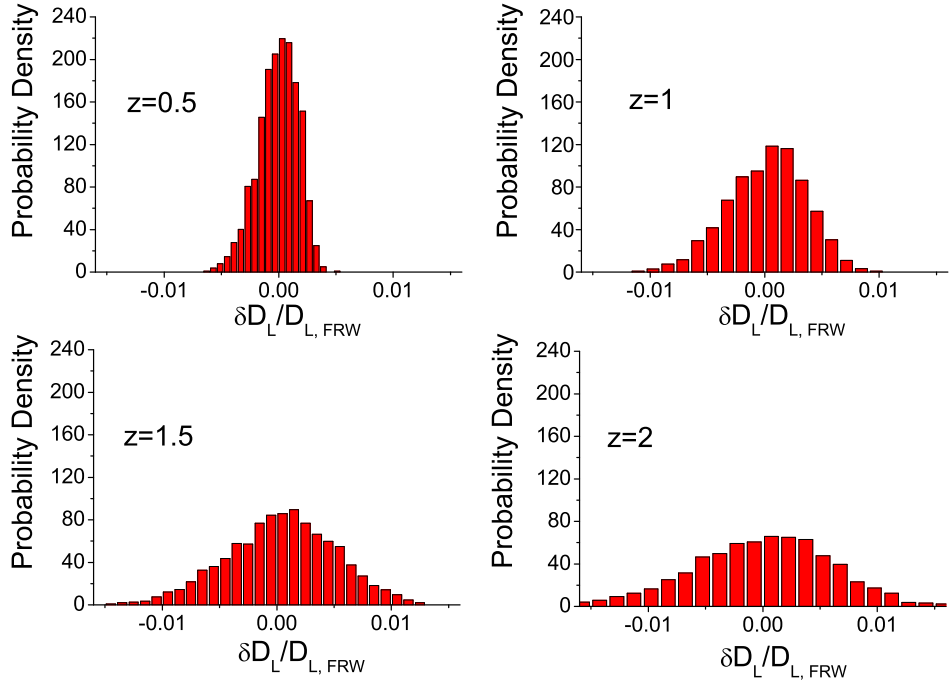}
 \caption{\it
The distribution of luminosity distances for various redshifts in the LTB Swiss-cheese
model if the inhomogeneities have a characteristic scale of $40\,h^{-1}$ Mpc.
}
 \label{fig5}
 \end{figure}

In this section we consider light beams that pass through several of the inhomogeneities
described in the previous sections.
The light is emitted at some time $\tb_s$ from a point
with $\rb=1$ at the edge of the inhomogeneous region. Its initial direction is
assumed to be random.
The initial
conditions for the beam area are given by eqs. (\ref{init1}), (\ref{init2}).
The light moves through the inhomogeneity and exits from a point with $\rb=1$.
Subsequently, the beam crosses the following inhomogeneity in a similar
fashion. The angle of entry into the new inhomogeneity is assumed to be
random again. The initial conditions are set by the values of
$\sqrt{\Ab}$ and $d\sqrt{\Ab}/{d\lb}$ at the end of the first crossing. This process
is repeated until the light arrives at the observer.
Of course, as time passes the profile of the inhomogeneities changes, as
depicted in fig. \ref{fig1}.

The assumption of entry into a spherical inhomogeneity at a random
azimuthal angle
is realized by selecting the impact factor $\ct_\phi$ with a
probability $\sim \ct_\phi d\ct_\phi$. The absence of the denominator
$\sqrt{1-\ct_\phi^2}$ that appears in the weight employed in the previous section
is justified by the fact that the source is located far from the center of the
inhomogeneity, so that $\ct_\phi \ll 1$.
This is not strictly true for the first 1 or 2 crossings, but the
induced error is small. In order
to place the observer outside the inhomogeneities we always
replace the final spherical region
crossed by the beam with a homogeneous configuration.

\begin{figure}[t]
\includegraphics[width=14cm, angle=0]{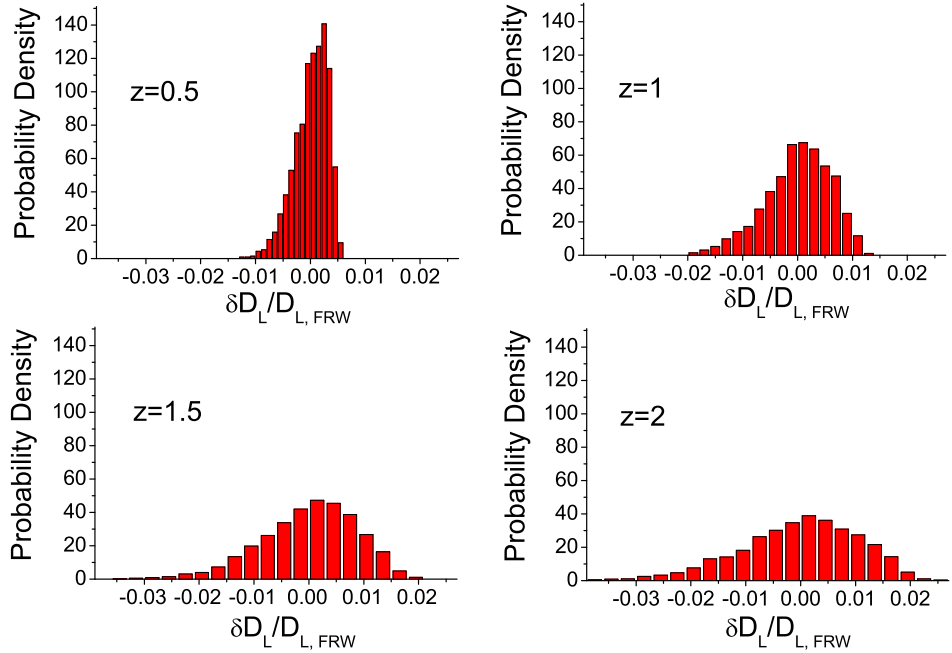}
 \caption{\it
Same as in fig. \ref{fig5} for a characteristic scale of $133\,h^{-1}$ Mpc.
}
 \label{fig6}
 \end{figure}

The essence of our procedure is that the light beam encounters
various inhomogeneities at various angles along its path. In this section we consider the
propagation of the beam only in the intervals $0\leq \rt \leq 1$ of coordinate systems
with origins at the center of the various inhomogeneities.
This means that we neglect the homogeneous region between the inhomogeneities that
we assumed in the previous region. (Essentially we assume that its width is
negligible.) The reason for this omission is that the
presence of the homogeneous region generates a bias towards negligible deviations of
the luminosity distance from its value in a homogeneous cosmology. For example,
if we consider light propagation in the intervals $0\leq \rt \leq 1.5$ as in the previous
section, a large number of beam trajectories propagate only within the
homogeneous regions with $1\leq \rt \leq 1.5$. These give a luminosity distance for
the source equal to that in the FRW cosmology. On the other hand, there are trajectories
that propagate through the inhomogeneities, for which the cancellation between positive
and negative contributions results in a negligible
total deviation of the luminosity distance from the value in a homogeneous background.
It is the latter events that we are interested in, while the former are rather unphysical.

The procedure we outlined above has an unsatisfactory element. The spherical
inhomogeneities are assumed to follow each other continuously along the beam
trajectory. If the angles of entry of the beam are non-zero, the resulting
geometry implies that there is an overlap of the inhomogeneities outside the
beam trajectory. This problem cannot be corrected as long as the assumption of spherical
symmetry of the inhomogeneities is maintained. A choice must be made between
an artificial bias in the luminosity distance if intermediate homogeneous regions are
introduced, or the overlap of the inhomogeneities outside the
beam trajectory. We have chosen the second
option, as we believe that it provides a more reliable estimate of the distribution
of luminosity distances.

\begin{figure}[t]
\includegraphics[width=14cm, angle=0]{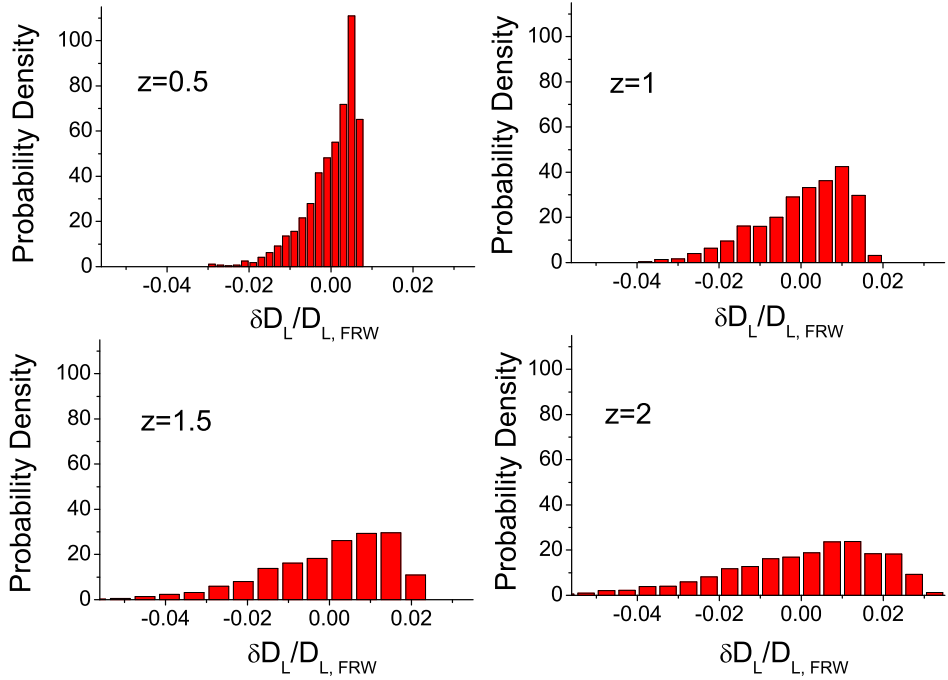}
 \caption{\it
Same as in fig. \ref{fig5} for a characteristic scale of $400\,h^{-1}$ Mpc.
}
 \label{fig7}
 \end{figure}

The total number of crossings determines the redshift
and the final beam area, related to the luminosity distance. We determine
the arrival time
at the observer by requiring that the redshift reach a specific value.
We repeat the calculation many times (at least 1000) for each value of the redshift and
plot the resulting distribution of luminosity distances.
The emission time is such that the
arrival time is $\tb_o\simeq 284$ for all the
redshifts that we consider. At this time the profile of the inhomogeneity is
very similar to the curve in fig. \ref{fig1} with the largest deviation from 1.
The deviations of the exact arrival time  $\tb_o$ from
the time in a homogeneous background is one order of magnitude smaller than the
respective deviation for the luminosity distance. This is in agreement with the discussion
in the previous section.

In figs. \ref{fig5}, \ref{fig6}, \ref{fig7} we depict the distributions of
the deviations of the luminosity distances from the value in
a homogeneous background
for various redshifts. The three figures correspond to
inhomogeneities with different characteristic length scales at the time
of the emission of light. The background through which the light propagates
is constructed as described in the previous sections. At the initial
time $\tb_i=0$ the inhomogeneities have the profile descibed in the previous section
with $\ex_1=-0.01$. The subsequent evolution is depicted in fig. \ref{fig1}. The
characteristic length scale of the inhomogeneities relative to the distance to the horizon
at the time $\tb_i=0$ is $r_0/(H^{-1}_i)=\Hb_i$. This quantity does not appear in
the rescaled evolution equations for the background. It appears only in the
rescaled geodesic equations. For this reason we can use the same background,
with an evolution depicted in fig. \ref{fig1}, in order to discuss inhomogeneities
of various length scales. The important phenomenological quantity is the
scale of the inhomogeneities today. This is given by
$R(t_f,r_0)/(H^{-1}_f)=\dot{R}(t_f,r_0)=\dot{\Rb}(\ttb_f,1)\Hb_i$. The rescaled present time
$\ttb_f$ is equal to the time of arrival of light signals to the observer $\ttb_o$
for all the cases we consider. As we mentioned already, $\ttb_f=\ttb_o\simeq 284$.
Our solution has $\dot{\Rb}(\tb_o)=0.133$.
Using $H^{-1}_f= 3\times 10^3\,h^{-1}$ Mpc we have
$R(t_f,r_0)=400\,h^{-1}\Hb_i$ Mpc.

The three figures \ref{fig5}, \ref{fig6}, \ref{fig7} correspond to
$\Hb_i=1/10$, 1/3, 1, respectively. They describe the effect on the propagation
of light of inhomogeneities with sizes 40, 133, $400\,h^{-1}$ Mpc today.
The present profile of the inhomogeneities has a density contrast
${\cal O} (1)$. Their
evolution, as modelled by the LTB metric, is
roughly consistent with the standard theory of structure growth.
The choices $\Hb_i=1/3$ and 1 lead to present-day inhomogeneities with length scales
larger than those in typical observations. The size of the same perturbations at
horizon crossing is larger than the value $\sim 10^{-5}$ implied by the
CMB. However, we have included them for two reasons. Firstly, because there are
indications that the presence of such large structures may be supported by
observations \cite{tomita,localvoid}.
Secondly, because we would like to understand if inhomogeneities with sizes comparable
to the horizon distance can have a significant effect on the luminosity distance for
a random location of the observer.

The total integral of
the distributions has been normalized to 1 in all cases, so that they are in fact
probability densities.
They have similar profiles that are asymmetric around
zero. Each distribution has a maximum at a value larger
than zero and a long tail towards negative values. The average deviation is zero to
a good approximation in all cases. This is expected according to our discussion of
flux conservation in the previous section. As long as the light propagation in
an inhomogeneous background does not modify significantly the redshift, the
energy may be redistributed in various directions, but the total flux is conserved and
remains the same as in a FRW background.

The longer tail of the distribution towards luminosity distances smaller than the one in
a homogeneous background is a consequence of the presence of a thin and dense spherical
shell around each central underdensity. The number of beam trajectories that
propagate through several shells is small. However, the focusing of the beam is
substantial for such beams and the resulting luminosity distance much shorter than
the average. The effect of the long tail is compensated by the shift of the
maximum of the distribution towards positive values.
The form of the distribution is very similar to that derived in studies modelling
the inhomogeneities through the standard Swiss-cheese model \cite{holz}.
In that case the strong focusing is generated by the very dense concentration of matter
at the center of each spherical inhomogeneity. We emphasize, however, that the two models
have a different region of applicability. The standard Swiss-cheese model is appropriate
for length scales of ${\cal O}(1)\, h^{-1}$ Mpc or smaller, while the LTB Swiss-cheese
model for
scales of ${\cal O}(10)\, h^{-1}$ Mpc or larger.

\section{Determination of cosmological parameters}
\setcounter{equation}{0}

\begin{figure}[t]
\includegraphics[width=12cm, angle=0]{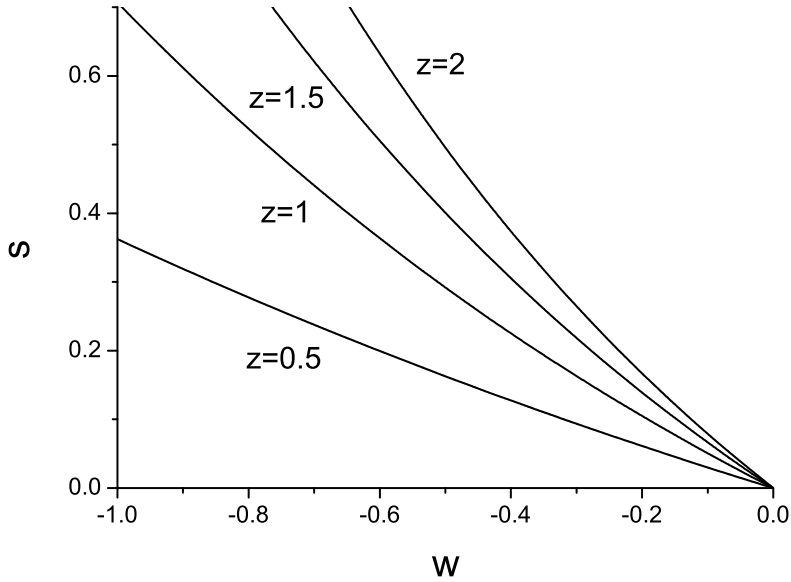}
 \caption{\it
The relative deviation of the luminosity
distance from its value for $w=0$ (non-relativistic matter) in homogeneous cosmology.
}
 \label{fig8}
 \end{figure}

The form of the distributions can be quantified in terms of two
parameters: The width of the distribution $\delta_d$ and the location of its maximum
$\delta_m>0$.
The first one characterizes the error induced to cosmological parameters
derived through the curve of the luminosity distance as a function of redshift,
while the second one the bias in such determinations. As discussed
extensively in ref. \cite{holz}, a small sample of data is expected to favour
values of the luminosity distance near the maximum of the distribution, and thus
generate a bias.
In figs. \ref{fig5}, \ref{fig6}, \ref{fig7} we observe that both
$\delta_d$ and $\delta_m$ grow with inreasing redshift $z$ and scale $\Hb_i$.
For $\Hb_i=1/10$ the average width $\delta_d$ increases from approximately 0.005 to
0.01 as $z$ increases from 0.5 to 2. The maximum is $\delta_m \lta 0.002$
for all $z$. The asymmetry of the distribution is very small.
For $\Hb_i=1/3$ the average width increases from approximately 0.005 to
0.02 as $z$ increases from 0.5 to 2. The maximum is $\delta_m \lta 0.002$ for all $z$.
The asymmetry of the distribution and the longer tail towards negative values are
clearly visible in fig. \ref{fig6}. For
$\Hb_i=1$ the average width increases from approximately 0.01 to
0.04 as $z$ increases from 0.5 to 2. The maximum increases from 0.005 to 0.01.
The asymmetry of the distribution is very distinctive in fig. \ref{fig7}.

The values of $\delta_d$ and $\delta_m$ are very small for all the values of
$\Hb_i$ that we considered. This implies that we do not expect a significant
effect on the cosmological parameters. As an interesting example we consider
the parameter $w$ that appears in the equation of state of the cosmological
fluid. In homogeneous cosmology the luminosity distance is
a function of $w$ and the redshift $z$. The relative deviation of the luminosity
distance from its value for $w=0$ (non-relativistic matter) is
\begin{equation}
s(w,z) = \frac{D_L(w,z)}{D_L(0,z)}-1
=\frac{1}{1+3w}\frac{1-\left(z+1\right)^{-(1+3w)/2}}{1-\left(z+1\right)^{-1/2}}-1.
\label{ddlwz} \end{equation}
We depict this function in fig. \ref{fig8}. For $w=-1$ the
luminosity distance is larger by roughly 35\% relative to $w=0$ for $z=0.5$.
For $z=1$ the relative increase is approximately 70\%.

\begin{figure}[t]
\includegraphics[width=12cm, angle=0]{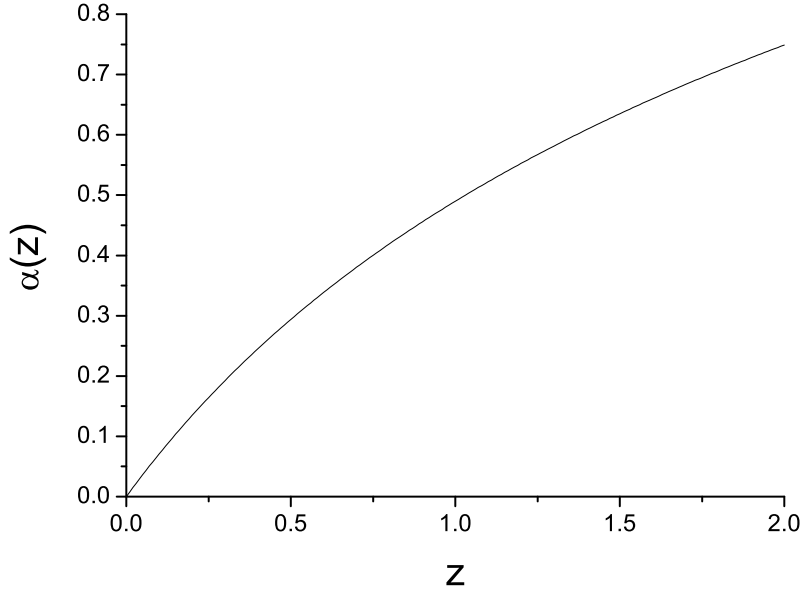}
 \caption{\it
The function $\alpha(z)=-\left.\partial s(w,z)/\partial w \right|_{w=0}$.
}
 \label{fig9}
 \end{figure}

The deviations of
$\delta_d(z)$ and $\delta_m(z)$ from zero because of the presence
of inhomogeneities can be attributed to deviations of $w$ from zero
if the cosmology is assumed to be homogeneous. This can be achieved by
identifying $\delta_d(z)$ or $\delta_m(z)$ with $s(w,z)$.
For small $w$ we have
\begin{equation}
s(w,z) \simeq -3\left(
1-\frac{1}{2}\frac{\ln(z+1)}{(z+1)^{1/2}-1}
\right) w =-\alpha(z) w.
\label{zw} \end{equation}
We depict the function $\alpha(z)$ in fig. \ref{fig9}.
For a given value of $z$ we can derive effective values of $w$ from the
relations $w_{eff}=-\delta_{d,m}(z)/\alpha(z)$.
It is clear from the values we quoted above for $\delta_d$,
$\delta_m$ and fig. \ref{fig9} that $|w_{eff}|\ll 1$.
As a result, for a random location of the observer
the perceived acceleration of the Universe cannot be attributed to
the modification of the luminosity distance by
large-scale inhomogeneities, even when their characteristic scale is smaller than
the horizon distance by less than a factor of 10.

On the other hand, the presence of inhomogeneities induces a
statistical error in the
value of $w$ deduced from astrophysical data, as well as a shift of its
average value if the sample is small.
According to our results, for $\Hb_i=1/10$ (and present size of the inhomogeneities
$40\,h^{-1}$ Mpc) the error
is $\dw\simeq 0.015$ for all $z$ between 0.5 and 2, while the average value $\bar{w}$
for a small sample is negative and of ${\cal O}(10^{-3})$.
For $\Hb_i=1/3$ (and present size of the inhomogeneities $133\,h^{-1}$ Mpc) 
the error increases from 0.015 to 0.025 as $z$ increases from
0.5 to 2, while the average value $\bar{w}$ is again negative and of ${\cal O}(10^{-3})$.
 For $\Hb_i=1$
(and present size of the inhomogeneities $400\,h^{-1}$ Mpc)
the error increases from 0.03 to 0.05, while the average
is $\wb \simeq -0.015$. The shift in the average value is always smaller than a
standard deviation.

The values of $\delta_d$ and $\delta_m$ can be compared to those generated by the
effects of gravitational
lensing at scales typical of galaxies or clusters of galaxies.
At such scales the Universe is modelled through the standard
Swiss-cheese model, with the mass of each inhomogeneity concentrated in a
very dense object at its center \cite{holz}.
The typical values of $\delta_d$ and $\delta_m$ are larger by at least an
order of magnitude than the ones we obtained.
The reason is that in our model the density contrast is always of ${\cal O}(1)$.
Our study indicates that the
effect on the luminosity distance grows when the length scale of
the inhomogeneities increases and becomes comparable to the horizon distance. However, the strong lensing effect of
a high concentration of mass, such as a galaxy or cluster of galaxies, gives
a much larger effect.

We conclude that within our model the presence of inhomogeneities with large length scales,
even comparable to the horizon
distance, and density contrast of ${\cal O}(1)$ does not influence significantly
the propagation of light if the source and the observer have random locations.
It is possible that our modelling of the Universe
is lacking some essential feature that could
generate a significant effect on the luminosity distance. For example, it has been
suggested that large fluctuations in the spatial
curvature may result in a strong backreaction on the
overall expansion \cite{buchert}. Unfortunately, there are no known exact models that
realize such a scenario. In their absence the effect on the luminosity distance cannot
be computed. In our model, there are variations of the local curvature which result
in the collapse of the overdense regions (either central overdensities, or shells
surrounding underdensities). The evolution is in approximate agreement with the
standard theory of structure formation. It seems difficult to reconcile
a much larger local curvature with a large-scale structure consistent with observations.
Another possibility is that the assumed spherical symmetry of the inhomogeneities in our modelling
of the Universe is too constraining. It is possible that photon propagation through 
inhomogeneities without such a symmetry results in a stronger modification of the 
luminosity distance. This point will be the subject of a future investigation.

\vspace {0.5cm}
\noindent{\bf Acknowledgments}\\
\noindent
This work was supported by the research program
``Pythagoras II'' (grant 70-03-7992)
of the Greek Ministry of National Education, partially funded by the
European Union.

\vskip 1.5cm

\newpage

\section{Appendix}

In this appendix we describe the solution of eq. (\ref{eq1}) through the use of perturbation
theory in $\Hb_i$. In this way we demonstrate that a spherical inhomogeneity induces a 
deviation of the luminosity distance
from its value in a homogeneous background which is of ${\cal O} (\Hb_i^2)$.   

The null constraint (\ref{gtb4}), keeping terms up to ${\cal O}(\sc)$, is 
${d\bt}/{d\brr}=\pm \sc, $
with the negative sign corresponding to ingoing
and the positive to outgoing geodesics.
We can set $t_s=0$ so the geodesic inside the inhomogeneity  is
\be \bt=-\sc(\brr-\brr_0)+{\cal O}(\sc^2) \ee 
for ingoing, 
and 
\be \bt=\sc(\brr+\brr_0)+{\cal O}(\sc^2) \ee
for outgoing geodesics.
We can treat $\bt$ as an ${\cal O}(\sc)$ quantity.

The initial configuration we use for this estimate has $\bar{\rho}(0,\brr)=0$ for
$\brr<\brr_1$ and $\bar{\rho}(0,\brr)=1/(1-\brr_1^3)$ for $\brr>\brr_1$.
From (\ref{eind}) we can calculate various derivatives of $\bR$ at
$\bt=0$:
\be \dot{\bR}'(\bt,\brr)=\dot{\bR}'(0,\brr)+\bt
\ddot{\bR}'(0,\brr)+{\cal O}(\sc^2), \ee 
\be \frac{\bR''}{\bR'}(\bt,\brr)=
\frac{\bt^2}{2}\ddot{\bR}''(0,\brr)+{\cal O}(\sc^3). \ee 
For $\brr>\brr_1$ we
have 
\bea \ddot{\bR}'(0,\brr)&=&\frac{r^3+2 \brr_1^3}{2 r^3
\left(\brr_1^3-1\right)} \\
\ddot{\bR}''(0,\brr)&=&-\frac{3 \brr_1^3}{r^4
\left(\brr_1^3-1\right)}. \eea 
For $\brr<\brr_1 $ both
$\ddot{\bR}'(0,\brr)$ and $\ddot{\bR}''(0,\brr)$ are zero.
For the initial configuration that we assume, $\ddot{\bR}$ is a continuous function of 
$\rb$. However, $\ddot{\bR}'$ is discontinuous at $\rb=\rb_1$ and $\rb=1$, while 
$\ddot{\bR}''$ has $\delta$-function singularities at the same points. 

The initial conditions for the solution of eq. (\ref{eq1}) for an ingoing beam 
can be taken $\sbA(1)=0$, $d\sbA(1) /d\brr=-1$, without loss of generality.
We use the expansion 
\be
\sbA=\sbAz+\sc \sbAf+\sc^2 \sbAs +{\cal O}(\Hb_i^3).
\ee
To zeroth order in $\sc$, eq. (\ref{eq1})
becomes ${d^2 \sbAz}/{d \brr^2}=0$, with solution
$\sbAz(\brr)=-(r-1)$ for ingoing and
$ \sbAz(\brr)=r+1$ for outgoing beams.
To first order in $\sc$, eq. (\ref{eq1})
gives ${d \sbAf}/{d \brr}=-2$, with solution
$\sbAf(\brr)=r^2-2r+1$ for ingoing and 
$\sbAf(\brr)=r^2+2r+1$ for outgoing beams.

To second order in $\sc$, and for $\brr>\brr_1$ and ingoing geodesics, we obtain
\bea \frac{d^2\sbAs}{d \brr^2}+\Biggl(2 \bt(\brr)\ddot{\bR}'(0,\brr)-
\frac{\bt^2}{2}\ddot{\bR}''(0,\brr)&+&\frac{\fb'(\brr)}{2} \Biggr) \frac{d
\sbAz}{d \brr}+2\frac{d
\sbAf}{d \brr}
\nonumber \\
&=&-\frac{3}{2} \bar{\rho}(0,\brr)  \sbAz.
\label{eqqq1} \eea
For $\brr<\brr_1$ and ingoing geodesics, we have 
\be \frac{d^2\sbAs}{d \brr^2}+\frac{\fb'(\brr)}{2}  \frac{d
\sbAz}{d \brr}+2\frac{d
\sbAf}{d \brr}=0.
\label{eqqq2} \ee
For $\brr<\brr_1$ and outgoing geodesics, we have
\be \frac{d^2\sbAs}{d \brr^2}+\frac{\fb'(\brr)}{2}  \frac{d
\sbAz}{d \brr}-2\frac{d
\sbAf}{d \brr}=0.
\label{eqqq3} \ee
Finally, for  $\brr>\brr_1$ and outgoing geodesics, we obtain 
\bea 
\frac{d^2\sbAs}{d \brr^2}+\Biggl(-2 \bt(\brr)\ddot{\bR}'(0,\brr)-
\frac{\bt^2}{2}\ddot{\bR}''(0,\brr)&+&\frac{\fb'(\brr)}{2} \Biggr) \frac{d
\sbAz}{d \brr}-2\frac{d
\sbAf}{d \brr}
\nonumber \\
&=&-\frac{3}{2} \bar{\rho}(0,\brr)  \sbAz.
\label{eqqq4} \eea
The above equations can be solved analytically through simple integration, with the 
values at the end of each interval determining the initial conditions for the
next one. The only non-trivial point is that the
$\delta$-function singularities of $\ddot{\bR}''$ at $\rb=\rb_1$ and $\rb=1$ 
induce discontinuities in the values of ${d\sbAs}/{d \brr}$ at these points. These must
be taken into account in a consistent calculation. The discontinuities can be
easily determined through the integration of eqs. (\ref{eqqq1})--(\ref{eqqq4}) in an infinitesimal
interval around each of these points. 
The remaining calculation is straightforward.
When the photon exits the inhomogeneity
at $\brr=1$ we
find that $\sqrt{\Ab}$ is given by eq. (\ref{final}).


\newpage

\vspace {0.5cm}
\noindent{\bf References}
\vspace {0.5cm}

\begin{thebibliography}{999}


\bibitem{accel1}
  A.~G.~Riess {\it et al.}  [Supernova Search Team Collaboration],
  Astron.\ J.\  {\bf 116} (1998) 1009
  [arXiv:astro-ph/9805201];
  Astrophys.\ J.\  {\bf 607} (2004) 665
  [arXiv:astro-ph/0402512];
\\
  S.~Perlmutter {\it et al.}  [Supernova Cosmology Project Collaboration],
  Astrophys.\ J.\  {\bf 517} (1999) 565
  [arXiv:astro-ph/9812133].

\bibitem{accel2}
  W.~J.~Percival {\it et al.}  [The 2dFGRS Collaboration],
  Mon.\ Not.\ Roy.\ Astron.\ Soc.\  {\bf 327} (2001) 1297
  [arXiv:astro-ph/0105252];
\\
  J.~L.~Sievers {\it et al.},
  Astrophys.\ J.\  {\bf 591} (2003) 599
  [arXiv:astro-ph/0205387].

\bibitem{wmap}
  D.~N.~Spergel {\it et al.}  [WMAP Collaboration],
  Astrophys.\ J.\ Suppl.\  {\bf 148} (2003) 175
  [arXiv:astro-ph/0302209].



\bibitem{rasanenn}
  S.~Rasanen,
  JCAP {\bf 0402} (2004) 003
  [arXiv:astro-ph/0311257];
\\
  E.~W.~Kolb, S.~Matarrese, A.~Notari and A.~Riotto,
  Phys.\ Rev.\ D {\bf 71} (2005) 023524
  [arXiv:hep-ph/0409038].


\bibitem{extra}
  T.~Futamase and M.~Sasaki,
  Phys.\ Rev.\  D {\bf 40} (1989) 2502.
\\
  M.~Kasai, T.~Futamase and F.~Takahara,
  Phys.\ Lett.\  A {\bf 147} (1990) 97.
\\
 F.~Hadrovic and J.~Binney,
  arXiv:astro-ph/9708110.
\\
  T.~Pyne and M.~Birkinshaw,
  Mon.\ Not.\ Roy.\ Astron.\ Soc.\  {\bf 348} (2004) 581
  [arXiv:astro-ph/0310841].

\bibitem{barausse}
  E.~W.~Kolb, S.~Matarrese, A.~Notari and A.~Riotto,
  Phys.\ Rev.\  D {\bf 71} (2005) 023524
  [arXiv:hep-ph/0409038].
\\
 E.~Barausse, S.~Matarrese and A.~Riotto,
  Phys.\ Rev.\  D {\bf 71} (2005) 063537
  [arXiv:astro-ph/0501152].


\bibitem{sachs}
  R.~K.~Sachs,
  Proc.\ Roy.\ Soc.\ London A {\bf 264} (1961) 309.


\bibitem{kantowski}
  R.~Kantowski,
  Astrophys.\ J.\  {\bf 155} (1969) 89.

\bibitem{swiss}
  A.~Einstein and E.~G.~Straus,
  Rev.\ Mod.\ Phys.\   {\bf 17} (1945) 120;
  \ {\it ibid.} \ {\bf 18} (1946) 148.

\bibitem{kantowski2}
  R.~Kantowski,
  Astrophys.\ J.\  {\bf 507} (1998) 483
  [arXiv:astro-ph/9802208];
  Phys.\ Rev.\ D {\bf 68} (2003) 123516
  [arXiv:astro-ph/0308419];
\\
  R.~Kantowski and R.~C.~Thomas,
  Astrophys.\ J.\ {\bf 561} (2001) 491
  [arXiv:astro-ph/0011176].



\bibitem{holz}
  D.~E.~Holz and R.~M.~Wald,
  Phys.\ Rev.\  D {\bf 58} (1998) 063501
  [arXiv:astro-ph/9708036];
\\
 D.~E.~Holz,
  Astrophys.\ J.\  {\bf 506}, L1 (1998)
  [arXiv:astro-ph/9806124];
\\
  D.~E.~Holz and E.~V.~Linder,
  Astrophys.\ J.\  {\bf 631}, 678 (2005)
  [arXiv:astro-ph/0412173].

\bibitem{sereno}
  M.~Sereno, G.~Covone, E.~Piedipalumbo and R.~de Ritis,
  Mon.\ Not.\ Roy.\ Astron.\ Soc.\  {\bf 327} (2001) 517
  [arXiv:astro-ph/0102486];
\\
  M.~Sereno, E.~Piedipalumbo and M.~V.~Sazhin,
  Mon.\ Not.\ Roy.\ Astron.\ Soc.\  {\bf 335} (2002) 1061
  [arXiv:astro-ph/0209181].



\bibitem{dyer}
C.~C.~Dyer and R.~C.~Roeder,
Astrophys.\ J.\  {\bf 180} (1973) L31;\ {\it ibid.}\ {\bf 189} (1974) 167





\bibitem{ltb}
G.~Lemaitre,
Gen. \ Rel. \ Grav. \ {\bf 29} (1997) 641;
\\
R.~C.~Tolman,
Proc.\ Nat.\ Acad.\ Sci.\  {\bf 20} (1934) 169;
\\
H.~Bondi,
Mon.\ Not.\ Roy.\ Astron.\ Soc.\  {\bf 107} (1947) 410.




\bibitem{mustapha}
 N.~Mustapha, C.~Hellaby and G.~F.~R.~Ellis,
  Mon.\ Not.\ Roy.\ Astron.\ Soc.\  {\bf 292} (1997) 817
  [arXiv:gr-qc/9808079].

\bibitem{celerier}
  M.~N.~Celerier,
  Astron.\ Astrophys.\  {\bf 353} (2000) 63
  [arXiv:astro-ph/9907206].

\bibitem{tbother}
  H.~Iguchi, T.~Nakamura and K.~i.~Nakao,
  Prog.\ Theor.\ Phys.\  {\bf 108} (2002) 809
  [arXiv:astro-ph/0112419];
\\
  K.~Bolejko,
  arXiv:astro-ph/0512103;
\\
  R.~Mansouri,
  arXiv:astro-ph/0512605;
\\
  R.~A.~Vanderveld, E.~E.~Flanagan and I.~Wasserman,
  Phys.\ Rev.\ D {\bf 74} (2006) 023506
  [arXiv:astro-ph/0602476];
\\
  D.~Garfinkle,
  Class.\ Quant.\ Grav.\  {\bf 23} (2006) 4811
  [arXiv:gr-qc/0605088];
\\
  D.~J.~H.~Chung and A.~E.~Romano,
  Phys.\ Rev.\ D {\bf 74} (2006) 103507
  [arXiv:astro-ph/0608403].



\bibitem{rasanen2}
  S.~Rasanen,
  JCAP {\bf 0411} (2004) 010
  [arXiv:gr-qc/0408097];
  JCAP {\bf 0611} (2006) 003
  [arXiv:astro-ph/0607626].


\bibitem{moffat}
  J.~W.~Moffat,
  JCAP {\bf 0510} (2005) 012
  [arXiv:astro-ph/0502110];
  arXiv:astro-ph/0505326.


\bibitem{alnes}
  H.~Alnes, M.~Amarzguioui and O.~Gron,
JCAP {\bf 0701} (2007) 007
  [arXiv:astro-ph/0506449];
  Phys.\ Rev.\ D {\bf 73} (2006) 083519
  [arXiv:astro-ph/0512006];
\\
 H.~Alnes and M.~Amarzguioui,
Phys.\ Rev.\  D {\bf 75} (2007) 023506
  [arXiv:astro-ph/0610331].

\bibitem{enqvist}
  K.~Enqvist and T.~Mattsson,
JCAP {\bf 0702} (2007) 019
  [arXiv:astro-ph/0609120].

\bibitem{chuang}
  C.~H.~Chuang, J.~A.~Gu and W.~Y.~Hwang,
  arXiv:astro-ph/0512651.


\bibitem{apo}
 P.~S.~Apostolopoulos, N.~Brouzakis, N.~Tetradis and E.~Tzavara,
  JCAP {\bf 0606} (2006) 009
  [arXiv:astro-ph/0603234].

\bibitem{biswas}
 T.~Biswas, R.~Mansouri and A.~Notari,
  arXiv:astro-ph/0606703.


\bibitem{tomita}
  K.~Tomita,
  Astrophys.\ J.\  {\bf 529}, 26 (2000);
  Astrophys.\ J.\  {\bf 529}, 38 (2000);
  Mon.\ Not.\ Roy.\ Astron.\ Soc.\  {\bf 326} (2001) 287
  [arXiv:astro-ph/0011484].


\bibitem{localvoid}
  W.~J.~Frith, G.~S.~Busswell, R.~Fong, N.~Metcalfe and T.~Shanks,
  Mon.\ Not.\ Roy.\ Astron.\ Soc.\  {\bf 345} (2003) 1049
  [arXiv:astro-ph/0302331].


\bibitem{brouzakis}
  N.~Brouzakis, N.~Tetradis and E.~Tzavara,
JCAP {\bf 0702} (2007) 013
  [arXiv:astro-ph/0612179].

\bibitem{lensing}
  P.~Schneider, J.~Ehlers and E.~E.~Falco,
  {\it Gravitational Lenses}, Springer-Verlag, Berlin.



\bibitem{biswas2}
  T.~Biswas and A.~Notari,
  arXiv:astro-ph/0702555.

\bibitem{israel}
W.~Israel, 
Nuovo Cim.\ B \textbf{44} (1966) 1. 



\bibitem{matching}
  H.~Sato,
  Prog.\ Theor.\ Phys.\  {\bf 76} (1986) 1250;
\\
  V.~A.~Berezin, V.~A.~Kuzmin and I.~I.~Tkachev,
  Phys.\ Rev.\  D {\bf 36} (1987) 2919;
\\
  S.~Khakshournia and R.~Mansouri,
  Phys.\ Rev.\  D {\bf 65} (2002) 027302
  [arXiv:gr-qc/0307023].


\bibitem{sphc1}
  J.~E.~Gunn and J.~R.~I.~Gott,
  Astrophys.\ J.\  {\bf 176}, 1 (1972);
\\
  A.~Cooray and R.~Sheth,
  Phys.\ Rept.\  {\bf 372} (2002) 1
  [arXiv:astro-ph/0206508].


\bibitem{partovi}
  M.~H.~Partovi and B.~Mashhoon,
  Astrophys.\ J.\  {\bf 276} (1984) 4.
\\
  N.~P.~Humphreys, R.~Maartens and D.~R.~Matravers,
  Astrophys.\ J.\  {\bf 477} (1997) 47
  [arXiv:astro-ph/9602033].



\bibitem{weinberg}
  S.~Weinberg,
  Astrophys.\ J.\  {\bf 208} (1976) L1

\bibitem{rose}
  H.~G.~Rose,
  Astrophys.\ J.\  {\bf 560} (2001) L15
  [arXiv:astro-ph/0106489].

\bibitem{snap}
  G.~Aldering, A.~G.~Kim, M.~Kowalski, E.~V.~Linder and S.~Perlmutter,
  Astropart.\ Phys.\  {\bf 27} (2007) 213
  [arXiv:astro-ph/0607030].

\bibitem{buchert}
  T.~Buchert,
  Class.\ Quant.\ Grav.\  {\bf 23} (2006) 817
  [arXiv:gr-qc/0509124];
\\
  T.~Buchert, J.~Larena and J.~M.~Alimi,
  Class.\ Quant.\ Grav.\  {\bf 23} (2006) 6379
  [arXiv:gr-qc/0606020].




\end{thebibliography}
\end{document}